\documentclass[onecolumn,superscriptaddress,showpacs,pra]{revtex4}
\usepackage{mathrsfs}
\usepackage{amssymb, amsbsy, amsmath, latexsym, dsfont, array, layout, graphicx,mathrsfs,color}

\newcommand{\ket}[1]{\left|{#1}\right\rangle}
\newcommand{\bra}[1]{\left\langle{#1}\right|}

\begin{document}

\title{Experimental investigation of the stronger uncertainty relations for all incompatible observables}
\author{Kunkun Wang}
\author{Xiang Zhan}
\affiliation{Department of Physics, Southeast University, Nanjing 211189, China}
\author{Zhihao Bian}
\affiliation{Department of Physics, Southeast University, Nanjing 211189, China}
\author{Jian Li}
\affiliation{Department of Physics, Southeast University, Nanjing 211189, China}
\author{Yongsheng Zhang\footnote{yshzhang@ustc.edu.cn}}
\affiliation{Key Laboratory of Quantum Information, University of Science and Technology of China, CAS, Hefei 230026, China}
\affiliation{Synergetic Innovation Center of Quantum Information and Quantum Physics, University of Science and Technology of China, Hefei 230026, China}
\author{Peng Xue\footnote{gnep.eux@gmail.com}}
\affiliation{Department of Physics, Southeast University, Nanjing 211189, China}

\begin{abstract}
The Heisenberg-Robertson uncertainty relation quantitatively expresses the impossibility of jointly sharp preparation of incompatible observables. However it does not capture the concept of incompatible observables because it can be trivial even for two incompatible observables. We experimentally demonstrate the new stronger uncertainty relations proposed by Maccone and Pati [Phys. Rev. Lett. {\bf 113}, 260401 (2014)] relating on that sum of variances are valid in a state-dependent manner and the lower bound is guaranteed to be nontrivial for two observables being incompatible on the state of the system being measured. The behaviour we find agrees with the predictions of quantum theory and obeys the new uncertainty relations even for the special states which trivialize Heisenberg-Robertson relation. We realize a direct measurement model and give the first experimental investigation of the strengthened relations.
\end{abstract}


\maketitle

\section{Introduction}

The famous uncertainty relation introduced by Werner Heisenberg is a basic feature of quantum theory enshrined in all textbooks~\cite{H27,WZ83}. The uncertainty principle dramatically illustrates the difference between classical and quantum mechanics. The principle bounds the uncertainties about the outcomes of two incompatible measurements, such as position and momentum on a particle. The more precisely the position of a particle is determined, the less precisely its momentum can be known, and vice versa. This lack of knowledge, so-called uncertainty, was quantified by Heisenberg using the standard deviation. If the measurement on a given particle is chosen from a set of two possible observables $A$ and $B$, the resulting bound on the uncertainty can be expressed in terms of the commutator,
\begin{equation}
\Delta A^2\Delta B^2\geq \left|\frac{1}{2}\langle\left[A,B\right]\rangle\right|^2,
\end{equation}
which is so-called Heisenberg-Robertson uncertainty relation~\cite{R29}.

Uncertainty relations are useful for a wide range of applications in quantum
technologies including quantum cryptography, quantum entanglement, quantum computation and general physics as well. In detail, they are useful for formulating quantum mechanics~\cite{H05,BTL07} (e.g. to justify the complex structure of the Hilbert space or as a fundamental building block for quantum mechanics and quantum gravity), for studying measurement-induced disturbance~\cite{O03,O04,BCC+10,BLW14}, for entanglement detection~\cite{BSLR03,G04,HBBB04}, for security analysis of quantum cryptography~\cite{FP96,RB09} and so on. Uncertainty relations were tested experimentally with neutronic~\cite{ESS+12,SSE+13,SSD+15} and photonic qubits~\cite{LXX+11,PHC+11,RDM+12,WHP+13,RBB+14,KBOE14}.

However Heisenberg-Robertson relation (1) does not fully capture the notion of incompatible observables since it is expressed in terms of the {\it product} of the variances of measurements of the observables. It means the product is zero when either of the two variances is zero even if the other variance is non-zero. This is the flaw in Heisenberg-Robertson relation.

To overcome this limitation, Maccone and Pati~\cite{MA14} have proposed two new uncertainty relations that employ the {\it sum} of the variances of measurements of general observables. Since the variance is a positive quantity, the sum will always be non-zero unless both variances are zero and this case only happens if the observables are ``compatible" meaning they have a definite value.

We report an experimental test of the new uncertainty relations for a single-photon measurement and demonstrate they are valid for states of a spin-$1$ particle~\cite{note}. Compared to the previous experiments which have come close to the original uncertainty limit~\cite{EMGM94,NAZ02,LBCC04,SAR+09,JAF+11}, but did not overcome the inherent flaw, our experimental results fully capture the notion of incompatible observables in contrast to Heisenberg-Robertson inequality thus making the uncertainty relations much stronger. The results also explain so-called complementarity---an extreme form of uncertainty (here variance is used as a measure of uncertainty), i.e., one of the two properties of a system is perfectly known, and the other is completely uncertain, which is a situation where Heisenberg-Robertson inequality fails to explain. Furthermore, in our experiment, every term can be obtained directly by the outcomes of the projective measurements. Our test realizes a direct measurement model which releases the requirement of quantum state tomography~\cite{LXX+11,PHC+11}.

\section{Theoretical framework}

Consider two observables $A$ and $B$ which are incompatible on the state $\ket{\psi}$. The new stronger uncertainty relations proposed by Maccone and Pati relating on that sum of variances are theoretically proven to be universally valid~\cite{MA14}. The first inequality~\cite{MA14} is
\begin{equation}
\Delta A^2+\Delta B^2\geq\pm i\langle\left[A,B\right]\rangle+\left|\bra{\psi}A\pm iB\ket{\psi^\perp}\right|^2,
\label{eq:first}
\end{equation}
where the sign should be chosen so that $\pm i\langle\left[A,B\right]\rangle$ (a real quantity) is positive. The inequality is valid for arbitrary state $\ket{\psi^\perp}$ orthogonal to $\ket{\psi}$. If the state $\ket{\psi}$ is not a joint eigenstate of $A$ and $B$, the lower bound of the inequality is nontrivial (nonzero) for almost any choice of $\ket{\psi^\perp}$. The uncertainty inequality in (2) is tight, i.e., it becomes an equality by maximizing over $\ket{\psi^\perp}$. For example, if $\ket{\psi}$ is one of the eigenstates of the observable $A$, the optimal choice of $\ket{\psi^\perp}$ is $\ket{\psi^\perp}_{B}=(B-\langle B\rangle)\ket{\psi}/\Delta B$, or $\ket{\psi^\perp}_{A}=(A-\langle A\rangle)\ket{\psi}/\Delta A$ for the case that $\ket{\psi}$ is one of the eigenstates of the observable $B$. If $\ket{\psi}$ is not an eigenstate of either, the optimal choice is $\ket{\psi^\perp}\propto (A\pm iB-\langle A\pm iB\rangle)\ket{\psi}$.

The second inequality with nontrivial bound~\cite{MA14} is
\begin{equation}
\Delta A^2+\Delta B^2\geq \frac{1}{2}\left|_{A+B}\bra{\psi^\perp}A+B\ket{\psi}\right|^2,
\label{eq:second}
\end{equation}
where $\ket{\psi^\perp}_{A+B}\propto (A+B-\langle A+B\rangle)\ket{\psi}$ is orthogonal to $\ket{\psi}$. The lower bound of (3) is nonzero unless $\ket{\psi}$ is an eigenstate of $A+B$. The inequality (3) becomes an equality if the state $\ket{\psi}$ is an eigenstate of $A-B$.

%The two inequalities can be combined in a single uncertainty relation relating on the sum of variances $\Delta A^2+\Delta B^2\geq \max \{\mathcal{L}_{(2)},\mathcal{L}_{(3)}\}$ with $\mathcal{L}_{(2),(3)}$ the right-hand side (RHS) of inequalities (2) and (3) respectively.

We show an example by choosing two components of the angular momentum for a spin-$1$ particle as two incompatible observables $A=J_x=\begin{pmatrix}
                                    0 & 1 & 0 \\
                                    1 & 0 & 1 \\
                                    0 & 1 & 0 \\
                                 \end{pmatrix}
$ and $B=J_y=\begin{pmatrix}
                 0 & i & 0 \\
                 -i & 0 & i \\
                 0 & -i & 0 \\
              \end{pmatrix}
$, and a family of states being measured~\cite{MA14}
\begin{equation}
\ket{\psi_\phi}=\sin\phi\ket{+}+\cos\phi\ket{-}=(\sin\phi,0,\cos\phi)^\text{T},
\label{eq:state}
\end{equation}
where $\ket{\pm}$ are the eigenstates of $J_z=\begin{pmatrix}
                 1 & 0 & 0 \\
                 0 & 0 & 0 \\
                 0 & 0 & -1 \\
\end{pmatrix}$ corresponding to the eigenvalues $\pm1$, and $\phi\in\left[0,\pi\right]$ is the coefficient. None of $\ket{\psi_\phi}$ is a joint eigenstate of $J_x$ and $J_y$, nor an eigenstate of $J_x+J_y$. The uncertainty relations in (2) and (3) are valid for the states. However for the special cases with $\phi=\pi/4$ and $\phi=3\pi/4$ Heisenberg-Robertson relation can be trivial because one of the measurement variances is zero.

Now we focus on the feasibility of implementation of measurements. The variances $\Delta J_{x(y)}=\langle J^2_{x(y)}\rangle-\langle J_{x(y)}\rangle^2$ can be calculated by the measured expectation values of $J_{x(y)}$ and $J^2_{x(y)}$. The first term of the right-hand side (RHS) of the inequality (2) can be calculated by the measured expectation value of $J_z=i\left[J_x,J_y\right]/2$. The second term can be rewritten as $\left|\bra{\psi_\phi}J_x\pm iJ_y\ket{\psi^\perp_\phi}\right|^2=\bra{\psi_\phi}C_\pm\ket{\psi_\phi}$, where \begin{equation}
C_\pm:=(J_x\pm iJ_y)\ket{\psi^\perp_\phi}\bra{\psi^\perp_\phi}(J_x\mp iJ_y)
\label{eq:C}
\end{equation}
are Hermitian operators. The second term of the RHS of (2) can be calculated by the expectation values of the operators $C_\pm$. The inequality (2) can be rewritten as $\Delta J_x^2+\Delta J_y^2\geq |\langle J_z\rangle|+\langle C_\pm\rangle$.

Similarly the inequality (3) can be rewritten as $\Delta J_x^2+\Delta J_y^2\geq \langle D\rangle$, where the observable of the measurement is
\begin{equation}
D:=\frac{1}{2}(J_x+J_y)\ket{\psi^\perp_\phi}_{J_x+J_y}\bra{\psi_\phi^\perp}(J_x+J_y).
\label{eq:D}
\end{equation}
The observables $C_{\pm}$ and $D$ are dependent on the choices of the states $\ket{\psi^\perp_\phi}$ and $\ket{\psi^\perp_\phi}_{J_x+J_y}$ respectively. Though the second inequality is tight too, for the family of states in (\ref{eq:state}) which are not the eigenstates of $J_x\pm J_y$, the inequality will not become an equality for any choice of $\ket{\psi^\perp_\phi}_{J_x+J_y}$.

\begin{figure}
   \includegraphics[width=.5\textwidth]{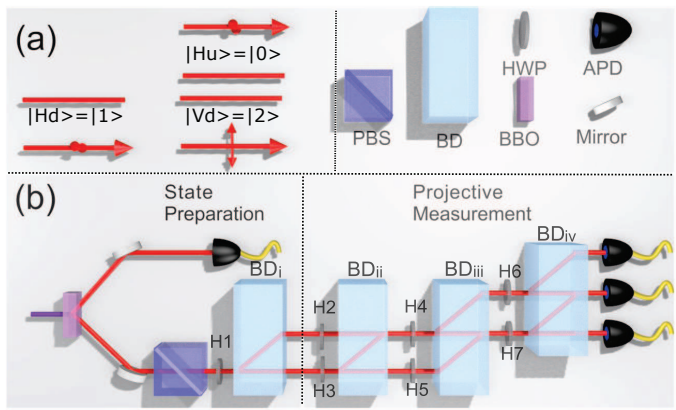}
   \caption{(a) Representation of qutrit. Here H and V denote the horizontal and vertical polarizations of the single photons, respectively. The subscripts u and d represent the upper and lower spatial modes of the single photons respectively. (b) Experimental setup. The herald single photons are created via type-I SPDC in a BBO crystal and are injected into the optical network. The PBS, HWP (H1) and BD$_{i}$ are used to generate a qutrit state $\ket{\psi_\phi}$. The HWPs (H2-H7) and three BDs are used to realize the projective measurements of observables $J_{x(y)}$, $J^2_{x(y)}$, $C_\pm$. To realize that of $D$, H4 and H5 are replaced by the QWPs (Q4 and Q5) at $45^\circ$.}
\label{setup}
\end{figure}

\section{Experimental implementation}

We report the experimental test of the new uncertainty relations for a single-photon measurement.

The experimental setup shown in Fig.~1 involves preparing the specific state (state preparation stage) and projective measurement on the system of interest (measurement stage). In the preparation stage, polarization-degenerate photon pairs at a wavelength of $801.6$nm are produced in a type-I spontaneous parametric down-conversion (SPDC) source using a $0.5$mm-thick $\beta$-barium-borate (BBO) nonlinear crystal, pumped by a CW diode laser with $90$mW of power~\cite{Xue1,Xue2,Xue3,Xue6}. The pump is filtered out with the help of an interference filter which restricts the photon bandwidth to $3$nm. With the detection of trigger photon the signal photon is heralded in the measurement setup. Experimentally this trigger-signal photon pair is registered by a coincidence count at two single-photon avalanche photodiodes (APDs) with $7$ns time window. Total coincidence counts are about $10^4$ over a collection time of $6$s.

A qutrit is represented by three
modes of the heralded single photons shown in Fig.~1(a) and the basis states $\ket{0}$, $\ket{1}$, and $\ket{2}$ are encoded
by the horizontal polarization of the photon in the upper
mode, the horizontal polarization of the photon in the lower mode, and the vertical polarization in the lower mode, respectively. The heralded single photons pass through a polarizing beam splitter (PBS) and a half-wave plate (HWP, H1) with the certain setting angle and then split by a birefringent calcite
beam displacer (BD) into two parallel spatial modes---upper and lower modes~\cite{Xue4,Xue5}.
The optical axis of the BD is cut so that vertically polarized light is directly transmitted and horizontal light undergoes a $3$mm lateral displacement into a neighboring mode. Thus the photons are prepared in the state $\ket{\psi_\phi}$ in Eq.~(\ref{eq:state}). We choose $\phi=j\pi/12$ ($j=1,...,12$), i.e., total twelve states for testing the uncertainty relation proposed in~\cite{MA14}. The setting angle $\theta_1(\phi)$ of H1 used for generating the state $\ket{\psi_\phi}$ satisfies $\theta_1(\phi)=\pi/4-\phi/2$.

\begin{figure}
   \includegraphics[width=.45\textwidth]{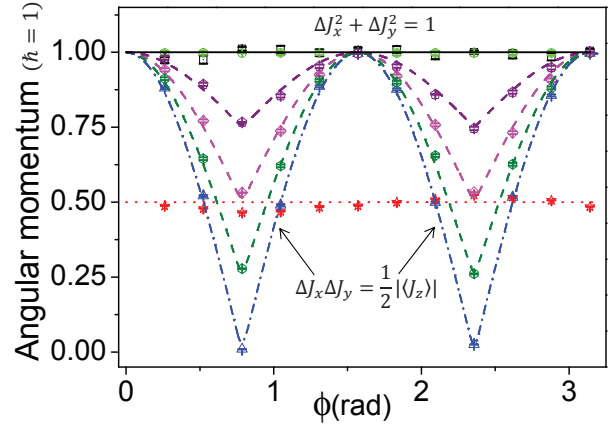}
   \caption{Experimental results.
   The solid black line corresponds to the LHS of the inequalities (2) and (3), i.e., $\Delta J_x^2+\Delta J_y^2=1$. The black squares represent the sum of the measured uncertainties of $\Delta J_x^2$ and $\Delta J_y^2$ with the twelve states $\ket{\psi_\phi}$. The green circles, olive hexagons, magenta diamonds and purple pentagons represent the experimental results of the RHS of inequality (2) with the optimal state $\ket{\psi^\perp_\phi}_\text{opt}$ and three randomly chosen states $\ket{\psi^\perp_\phi}_\text{1,2,3}$ for each of the twelve values of $\phi$. The red dotted lines corresponds to the bound of the inequality (3) and red stars represent the measured $\langle D\rangle$ for the twelve states $\ket{\psi^\perp_\phi}_{J_x+J_y}$. The blue dotted-dashed curves and triangles represent the theoretical predictions and experimental results of the product of the uncertainties and the expectation value of the commutator (Heisenberg-Robertson relation). Error bars indicate the statistical uncertainty.
   }
\label{setup}
\end{figure}

The left-hand side (LHS) of the inequalities (2) and (3), i.e., the upper bound is $\Delta J_x^2+\Delta J_y^2=1$, which is constant for all the twelve states being measured. For half of the states with $\phi=\pi/12,\pi/6,\pi/4,5\pi/6,11\pi/12,\pi$, because of $i\langle\left[J_x,J_y\right]\rangle\leq0$, the RHS of inequality (2) can be rewritten as $2|\langle J_z\rangle|+\langle C_-\rangle$. Whereas, for the rest six states being measured, it is $2|\langle J_z\rangle|+\langle C_+\rangle$.

In the measurement stage, cascaded interferometers which consist of BDs and wave plates (WPs) are used to implement an operation $U=\sum_i\ket{i}\bra{m_i}$, where $\ket{m_i}$ ($i=0,1,2$) is the eigenstate of the observable $M=\sum_i m_i\ket{m_i}\bra{m_i}$ according to the eigenvalue $m_i$. The BD$_{iv}$ is used to map the basis states of qutrit to three spatial modes and to accomplish the projective measurement $\{\ket{m_0}\bra{m_0},\ket{m_1}\bra{m_1},\ket{m_2}\bra{m_2}\}$ of the observable $M$ on a qutrit state $\ket{\psi_\phi}$ along with APDs. The outcomes give the measured probability $p_{m_i}=\left|\bra{\psi_\phi}m_i\rangle\right|^2$, which equals to the probability $p_{i}=\text{Tr}(\ket{\psi_\phi}\bra{\psi_\phi}U^\dagger\ket{i}\bra{i}U)$ of the photons being measured in the state $\ket{i}$. The expected value of the observable $M$ can be calculated by the measured probabilities and the eigenvalues as $\langle M\rangle=\sum_i m_ip_{m_i}$. Similarly one can calculate the variance $\Delta M$ of $M$ with the outcomes of the projective measurements on the state $\ket{\psi_\phi}$.

To test the inequalities, we need to measure the observables $J_{x(y)}$, $J^2_{x(y)}$, $J_z$, $C_\pm$ in (\ref{eq:C}) and $D$ in (\ref{eq:D})~\cite{supp}. The unitary operation which performs a projective measurement is $SU(3)$, which can be realized by three substeps. Each of the substeps applies a rotation on two of the basis states $\{\ket{0},\ket{1},\ket{2}\}$ and keeps the rest one unchanged. Each of the substeps can be realized by two HWPs and BD. One of the HWPs is used to rotating the qutrit, the other is used to compensate the optical delay, and the BD is used to split the photons with different polarizations into different modes. The setting angles of H3, H4, H7 (or H6) can be calculated by the parameters of the projective measurement, the setting angles of H2, H5 are chosen to be $45^\circ$, and that of H6 is $0^\circ$ (or that of H7 is $-45^\circ$) to compensate the optical delay. The photons are detected by APDs, in coincidence with the trigger photons. The probabilities $p_i$ ($i=0,1,2$) are obtained by normalizing photon counts in the $i$th spatial mode to total photon counts.

To test the first inequality (2), for the twelve states $\ket{\psi_\phi}$ we choose, the optimization of $\ket{\psi_\phi^\perp}$ (namely, the choice that maximizes the lower bound and saturates the inequality) is independent of $\phi$ and takes the form \begin{equation}
\ket{\psi_\phi^\perp}_\text{opt}=(0,1,0)^\text{T}.
\end{equation}
Then for each $\phi$, we randomly choose three states
\begin{align}
&\ket{\psi_\phi^\perp}_{1}=\frac{\sqrt{3}}{2}(\cos\phi,\frac{\sqrt{3}}{3},-\sin\phi)^\text{T},\nonumber\\
&\ket{\psi_\phi^\perp}_{2}=\frac{\sqrt{2}}{2}(\cos\phi,1,-\sin\phi)^\text{T},\nonumber\\ &\ket{\psi_\phi^\perp}_{3}=\frac{1}{2}(\cos\phi,\sqrt{3},-\sin\phi)^\text{T}
\end{align}
which are orthogonal to $\ket{\psi_\phi}$ to test the inequality (2). For the optimal choice of the orthogonal state $\ket{\psi^\perp_\phi}_\text{opt}$, as well as the three others $\ket{\psi^\perp_\phi}_{1,2,3}$ , the projective measurement of observable $C_\pm$ can be realized by tuning the setting angles of the HWPs (H2-H7)~\cite{supp}.

Similarly, to test the second inequality (3), the measurement of the observable $D$ in (\ref{eq:D}) can be implemented by the setup in Fig.~1(b) by replacing the HWPs (H4 and H5) by the quarter-wave plates (QWPs, Q4 and Q5) respectively and setting angles $45^\circ$ for all the twelve states $\ket{\psi_\phi}$.

In Fig.~2, we show the direct demonstration of the two new uncertainty relations in (2) and (3). The LHS of inequalities (2) and (3), i.e., sum of the uncertainties $\Delta J^2_x+\Delta J^2_y$ is constant for the family of states in (\ref{eq:state}) as $\Delta J^2_x+\Delta J^2_y=1$. The experimental results of the LHS of inequalities are calculated from the measured data of observables $J_{x(y)}$ and $J^2_{x(y)}$ and fit the theoretical predictions well. The experimental results of the RHS of inequality (2) with the optimal choice of the state $\ket{\psi^\perp_\phi}_\text{opt}$ and the states $\ket{\psi^\perp_\phi}_{1,2,3}$ which are chosen randomly are shown with different symbols. It is clear that the bound (2) is always satisfied and outperforms Heisenberg-Robertson relation for arbitrary states $\ket{\psi^\perp_\phi}$ orthogonal to $\ket{\psi_\phi}$. All data of the RSH are above the lower curves which are the product of the uncertainties
and the expectation value of the commutator. Thus the new uncertainty relation in (2) is more strengthened compared to Heisenberg-Robertson relation. For the optimal choice $\ket{\psi^\perp_\phi}_\text{opt}$ which is independent of $\phi$, the experimental results fit the upper bound well. Thus the inequality becomes an equality, which shows the new uncertainty inequality (2) is tight.

For the inequality (3), due to the orthogonal state
\begin{equation}
\ket{\psi^\perp_\phi}_{J_x+J_y}=(0,1,0)^\text{T}
\end{equation}
which is independent of $\phi$, the theoretical prediction of the RHS is constant as $\langle D\rangle=0.5$, which fits our data well and satisfies the uncertainty inequality.

Our experimental results show that for the state $\ket{\psi_\phi}$ is an eigenstate of one of the two observables (in our case, $\phi=\pi/4$ and $\phi=3\pi/4$) which trivializes Heisenberg-Robertson relation, the lower bound of the new uncertainty inequalities is always nontrivial unless $\ket{\psi_\phi}$ is a joint eigenstate of the two observables.

Though as the previous experiments~\cite{LXX+11,PHC+11}, both sides of the inequalities (2) and (3) can be calculated from the density matrices of $\ket{\psi_\phi}$ which are characterized by quantum state tomography. In our experiment, every term of inequalities can be obtained directly by the outcomes of the projective measurements, and the experimental results are in a good agreement with theoretical predictions. Our test realizes a direct measurement model which much simplifies the experimental realization and releases the requirement of
quantum state tomography. It is much more ``user-friendly" compared to those require reconstruction of the state being measured by carrying out a set of measurements through tomographic means and calculated the expected values of the
measurement of observables.

Furthermore, our technique can be used to realize arbitrary $SU(3)$ unitary operation and arbitrary projective
measurements of a qutrit. An arbitrary $SU(3)$ unitary operation on qutrit can be decomposed into three matrices which can be realized by a transformation on two modes of the qutrit and keeping the third mode not affected~\cite{AACB13,RSG99}. Conveniently, two-mode transformations can then be implemented using WPs acting on the two polarization modes propagating in the same spatial mode. Thus we are able to apply transformations to any pair of modes.

\section{Conclusion}

We have demonstrated a method for experimentally testing the new uncertainty relations. This has allowed us to test the two uncertainty inequalities. Our experimental results clearly illustrate the new uncertainty relations between two components of the angular momentum. Our demonstration is the first evidence for the validity of the new relations proposed to be
universally valid. Our work conclusively shows that the new uncertainty relations are stronger and general compared to Heisenberg-Robertson uncertainty relation. The experimental results confirm that even for the special states which trivialize Heisenberg-Robertson relation, the uncertainties of the two observables obey the new relations, and shed light on fundamental limitations of quantum measurement. A correct understanding and experimental confirmation of a fundamental limitation of measurements will not only foster insight into foundational problems but also advance the precision measurement technology in quantum information processing, for instance on the debate the standard quantum limit for monitoring free-mass position~\cite{Y83,O88,M88}.

\appendix

\section{verify the new uncertainty relations with a three-level system}

The photonic states $\ket{\psi_\phi}$ can be generated in a type-I spontaneous parametric down-conversion (SPDC) process. For different $\phi$, one can vary the setting angle of the half-wave plate (HWP, H1). The angles of the HWPs for state preparation are listed in Table~\ref{table:1}. The density matrix
of the initial state $\ket{\psi_\phi}$ is characterized by the quantum state tomography process with nine measurement settings and the average fidelity for the twelve states is more than $0.988$.%, in which the LHS and RSH of the two new uncertainty inequalities can be calculated.
%In the basis $\{\ket{0},\ket{1},\ket{2}\}$, the HWP takes the form
%\begin{equation}
%R_\text{HWP}=\begin{pmatrix}
%                 \cos2\theta & \sin2\theta \\
%                 \sin2\theta & -\cos2\theta  \\
%\end{pmatrix}
%\end{equation}
%with the setting angle $\theta$ and applies a polarization rotation.

\begin{table*}[htbp]
%\scriptsize
\caption{The setting angles of the HWP (H1) for the state preparation stage.}
\begin{tabular}{c||c|c|c|c|c|c|c|c|c|c|c|c}
\hline
$\phi$(rad) & $\frac{\pi}{12}$ & $\frac{\pi}{6}$ & $\frac{\pi}{4}$ & $\frac{\pi}{3}$ & $\frac{5\pi}{12}$ & $\frac{\pi}{2}$& $\frac{7\pi}{12}$ & $\frac{2\pi}{3}$ & $\frac{3\pi}{4}$ & $\frac{5\pi}{6}$ &$\frac{11\pi}{12}$&$\pi$\\
\hline\hline
H1 & $37.50^\circ$ & $30.00^\circ$ & $22.50^\circ$ & $15.00^\circ$ & $7.50^\circ$ & $0.00^\circ$ & $-7.50^\circ$ & $-15.00^\circ$ & $-22.50^\circ$ & $-30.00^\circ$ & $-37.50^\circ$ & $-45.00^\circ$\\ \hline
%Fidelity & $0.9911$ & $0.9897$ & $0.9849$ & $0.9881$ & $0.9894$ & $0.9945$ & $0.9937$ & $0.9831$ & $0.9929$ & $0.9838$ & $0.9965$ & $0.9856$\\ \hline
\end{tabular}
\label{table:1}
\end{table*}

In the measurement stage, cascaded interferometers which consist of beam displacers (BDs) and wave plates (WPs) are used to implement projective measurements of the observables $J_{x(y)}$, $J_z$, $J^2_{x(y)}$, $C_{\pm}$ and $D$. The angles of the HWPs (H2-H7) for state preparation are listed in Table~\ref{table:2}. For some observables, we replace the HWPs (H4 and H5) by quarter-wave plates (QWPs, Q4 and Q5) with the setting angles $45^\circ$. The polarization analysis measurement setup containing
QWPs, HWPs and BDs can be used to perform measurements of the corresponding observable on photons.
% which takes the form \begin{equation}
%R_\text{QWP}(\vartheta)=\begin{pmatrix}
%                 \cos^2\vartheta+i\sin^2\vartheta & (1-i)\sin\vartheta\cos\vartheta \\
%                 (1-i)\sin\vartheta\cos\vartheta & \sin^2\vartheta+i\cos^2\vartheta  \\
%\end{pmatrix}
%\end{equation}
%and applies a polarization rotation.

The technique of direct state transformations with optical elements we use here can realize arbitrary $SU(3)$ unitary operation and arbitrary projective measurements of a qutrit. An arbitrary $SU(3)$ unitary operation on qutrit can be decomposed into three matrices which can be realized by a transformation on two modes of the qutrit and keeping the third mode not affected. Conveniently, two-mode transformations can then be implemented using WPs acting on the two polarization modes propagating in the same spatial mode. Thus we are able to apply transformations to any pair of modes.

Now we show an example on how to realize the measurement of observable $J_x$ via WPs and BDs and the state transformations of these optical elements.

The
measurement of observable $J_x$ can be realized via six half wave plates and three
BDs. The unitary operation which performs the measurement on the qutrit can be written as
\begin{equation}U=\begin{pmatrix}
                                    \frac{1}{2} & -\frac{1}{\sqrt{2}} & \frac{1}{2} \\
                                    -\frac{1}{\sqrt{2}} & 0 & \frac{1}{\sqrt{2}} \\
                                    \frac{1}{2} & \frac{1}{\sqrt{2}} & \frac{1}{2} \\
                                 \end{pmatrix}=U_3U_2U_1,
\end{equation}
where we have \begin{equation}U_1=\begin{pmatrix}
                                    1 & 0 & 0 \\
                                    0 & \sqrt{\frac{2}{3}} & -\frac{1}{\sqrt{3}} \\
                                    0 & -\frac{1}{\sqrt{3}} & -\sqrt{\frac{2}{3}} \\
                                 \end{pmatrix},  U_2=\begin{pmatrix}
                                    \frac{1}{2} & -\frac{\sqrt{3}}{2} & 0 \\
                                    \frac{\sqrt{3}}{2} & \frac{1}{2} & 0 \\
                                    0 & 0 & 1 \\
                                 \end{pmatrix}, U_3=\begin{pmatrix}
                                    1 & 0 & 0 \\
                                    0 & -\sqrt{\frac{2}{3}} & -\frac{1}{\sqrt{3}} \\
                                    0 & \frac{1}{\sqrt{3}} & -\sqrt{\frac{2}{3}} \\
                                 \end{pmatrix}.
\end{equation}
The three unitary operations $U_i$ ($i=1,2,3$) can be implemented by a single-qubit
rotation on two of three modes and keeping the other one unchanged.

The HWP (H3) at $-17.63^\circ$ is applied on the lower mode and implements a rotation
$\begin{pmatrix}
\sqrt{\frac{2}{3}} & -\frac{1}{\sqrt{3}}\\
-\frac{1}{\sqrt{3}} & -\sqrt{\frac{2}{3}}\\
\end{pmatrix}$ on the polarizations of photons in this mode, while keeps the
polarizations of the photons in the upper mode unchanged. Thus $U_1$ is realized.
The HWP (H2) at $45^\circ$ changes the polarizations of the photons in the upper mode
from horizontal to vertical and after BD$_{ii}$ the vertically
polarized photons go straight and are still in the upper mode. Whereas, the horizontally polarized photons in the lower mode go up into the upper mode. Then we can
realize a rotation $\begin{pmatrix}
-\frac{\sqrt{3}}{2} & \frac{1}{2}\\
\frac{1}{2} & \frac{\sqrt{3}}{2}\\
\end{pmatrix}$ on the polarizations of photons in the upper mode via H4 at $75^\circ$
and keep the polarizations of photons in the lower mode unchanged. Thus $U_2$ is
realized. Similarly, we use H5 and BD$_{iii}$ to move the photons with
different polarization into the lower mode and use H7 at $-62.63^\circ$ to realize
a rotation $\begin{pmatrix}
-\frac{1}{\sqrt{3}} & -\sqrt{\frac{2}{3}}\\
-\sqrt{\frac{2}{3}} & \frac{1}{\sqrt{3}}\\
\end{pmatrix}$ and keep the polarization of photons in the upper mode unchanged.
The HWP (H6) at $0$ is used as optical compensator. Thus $U_3$ is realized.

The basis states of qutrit $\ket{0}$, $\ket{1}$, and $\ket{2}$ are encoded
by the horizontal polarization of the photon in the upper
mode $\ket{Hu}$, the horizontal polarization of the photon in the lower mode
$\ket{Hd}$, and the vertical polarization in the lower mode $\ket{Vd}$, respectively.
The initial state \begin{equation}\ket{\phi}=\sin\phi\ket{Hu}+\cos\phi\ket{Vd}\end{equation} becomes
\begin{equation}\sin\phi\ket{Hu}-\frac{1}{\sqrt{3}}\cos\phi\ket{Hd}-\sqrt{\frac{2}{3}}\cos\phi\ket{Vd}\end{equation}
after H3 is applied on the polarizations of photons in the lower mode. After H2 is
applied, it becomes \begin{equation}
\sin\phi\ket{Vu}-\frac{1}{\sqrt{3}}\cos\phi\ket{Hd}-\sqrt{\frac{2}{3}}\cos\phi\ket{Vd}.
\end{equation}
Going through the following BD (BD$_{ii}$), the state is
\begin{equation}
-\frac{1}{\sqrt{3}}\cos\phi\ket{Hu}+\sin\phi\ket{Vu}-\sqrt{\frac{2}{3}}\cos\phi\ket{Vd}.
\end{equation}
The HWP (H4) is applied on the polarizations of photons in the upper mode and the
state becomes
\begin{equation}
\frac{1}{2}(\cos\phi+\sin\phi)\ket{Hu}+(-\frac{1}{2\sqrt{3}}\cos\phi+\frac{\sqrt{3}}{2}\sin\phi)
\ket{Vu}-\sqrt{\frac{2}{3}}\cos\phi\ket{Vd}.
\end{equation}
After H5 is applied on the photons in the lower mode,
$\ket{Vd}$ is rotated to $\ket{Hd}$. Going through the following BD$_{iii}$,
the state is \begin{equation}
\frac{1}{2}(\cos\phi+\sin\phi)\ket{Hu}-\sqrt{\frac{2}{3}}\cos\phi\ket{Hd}+(-\frac{1}{2\sqrt{3}}\cos\phi+\frac{\sqrt{3}}{2}\sin\phi)
\ket{Vd}.
\end{equation}
The HWP (H7) is applied on the polarizations of photons in
the lower mode and the state becomes
\begin{equation}
\frac{1}{2}(\cos\phi+\sin\phi)\ket{Hu}
+\frac{1}{\sqrt{2}}(\cos\phi-\sin\phi)\ket{Hd}+\frac{1}{2}(\cos\phi+\sin\phi)\ket{Vd}
\end{equation}
which is as same as $U\ket{\phi}$.

\begin{table}[htbp]
%\scriptsize
\caption{The setting angles of the HWPs (or QWP) for the projective measurement stage. Here ``$-$" denotes the corresponding WP is removed from the optical circuit.}
\begin{tabular}{c||c|c|c|c|c|c|c|c}
\hline
Observable & H2& H3 & H4 & Q4 & H5 & Q5 & H6& H7 \\
\hline\hline
$J_x$ & $45.00^\circ$ &$-17.63^\circ$ & $75.00^\circ$ & $-$ & $45.00^\circ$& $-$ &$0.00^\circ$& $-62.63^\circ$ \\ \hline
$J^2_x$ &$45.00^\circ$ & $90.00^\circ$ & $0.00^\circ$ & $-$ & $45.00^\circ$& $-$ &$0.00^\circ$& $22.50^\circ$ \\ \hline
$J_y$ & $45.00^\circ$ &$17.63^\circ$ & $-15.00^\circ$ & $-$ & $45.00^\circ$& $-$ &$0.00^\circ$& $-62.63^\circ$ \\ \hline
$J^2_y$ & $45.00^\circ$ &$90.00^\circ$ & $90.00^\circ$ & $-$ & $45.00^\circ$& $-$ &$0.00^\circ$& $22.50^\circ$ \\ \hline
$J_z$ & $45.00^\circ$ &$0.00^\circ$ & $45.00^\circ$ & $-$ & $45.00^\circ$& $-$ &$0.00^\circ$& $-45.00^\circ$ \\ \hline\hline
$C_\pm(\ket{\psi^\perp_\phi}_\text{opt})$ & $45.00^\circ$ &$0.00^\circ$ & $45.00^\circ$ & $-$ & $45.00^\circ$ & $-$&$0.00^\circ$& $-45.00^\circ$ \\ \hline
$C_\pm(\ket{\psi^\perp_{\pi/12}}_1)$ & $45.00^\circ$ &$0.00^\circ$ & $45.00^\circ$ & $-$& $45.00^\circ$ & $-$&$-32.93^\circ$& $-45.00^\circ$ \\ \hline
$C_\pm(\ket{\psi^\perp_{\pi/12}}_2)$  &$45.00^\circ$ & $0.00^\circ$ & $45.00^\circ$ & $-$& $45.00^\circ$ & $-$&$-37.74^\circ$& $-45.00^\circ$ \\ \hline
$C_\pm(\ket{\psi^\perp_{\pi/12}}_3)$  & $45.00^\circ$ &$0.00^\circ$ & $45.00^\circ$ & $-$& $45.00^\circ$  & $-$ &$-40.75^\circ$& $-45.00^\circ$ \\ \hline
$C_\pm(\ket{\psi^\perp_{\pi/6}}_1)$  & $45.00^\circ$ &$0.00^\circ$ & $45.00^\circ$ & $-$& $45.00^\circ$ & $-$ &$-24.55^\circ$& $-45.00^\circ$ \\ \hline
$C_\pm(\ket{\psi^\perp_{\pi/6}}_2)$  & $45.00^\circ$ &$0.00^\circ$ &  $45.00^\circ$ & $-$& $45.00^\circ$ & $-$ &$-31.72^\circ$& $-45.00^\circ$ \\ \hline
$C_\pm(\ket{\psi^\perp_{\pi/6}}_3)$  & $45.00^\circ$ &$0.00^\circ$ & $45.00^\circ$ & $-$& $45.00^\circ$ & $-$ &$-36.95^\circ$& $-45.00^\circ$ \\ \hline
$C_\pm(\ket{\psi^\perp_{\pi/4}}_1)$  & $45.00^\circ$ & $0.00^\circ$& $45.00^\circ$ & $-$& $45.00^\circ$ & $-$ &$-19.62^\circ$& $-45.00^\circ$ \\ \hline
$C_\pm(\ket{\psi^\perp_{\pi/4}}_2)$  & $45.00^\circ$ & $0.00^\circ$& $45.00^\circ$ & $-$& $45.00^\circ$ & $-$ &$-27.37^\circ$& $-45.00^\circ$ \\ \hline
$C_\pm(\ket{\psi^\perp_{\pi/4}}_3)$  & $45.00^\circ$ & $0.00^\circ$ & $45.00^\circ$ & $-$& $45.00^\circ$ & $-$ &$-33.90^\circ$& $-45.00^\circ$ \\ \hline
$C_\pm(\ket{\psi^\perp_{\pi/3}}_1)$  & $45.00^\circ$ & $0.00^\circ$ & $45.00^\circ$ & $-$& $45.00^\circ$ & $-$ &$0.00^\circ$& $-69.55^\circ$ \\ \hline
$C_\pm(\ket{\psi^\perp_{\pi/3}}_2)$  & $45.00^\circ$ & $0.00^\circ$ & $45.00^\circ$ & $-$ & $45.00^\circ$ & $-$ &$0.00^\circ$& $-76.72^\circ$ \\ \hline
$C_\pm(\ket{\psi^\perp_{\pi/3}}_3)$  & $45.00^\circ$ & $0.00^\circ$ & $45.00^\circ$ & $-$ & $45.00^\circ$ & $-$ &$0.00^\circ$& $-81.95^\circ$ \\ \hline
$C_\pm(\ket{\psi^\perp_{5\pi/12}}_1)$  & $45.00^\circ$ & $0.00^\circ$& $45.00^\circ$ & $-$ & $45.00^\circ$ & $-$ &$0.00^\circ$& $-77.92^\circ$ \\ \hline
$C_\pm(\ket{\psi^\perp_{5\pi/12}}_2)$  & $45.00^\circ$ & $0.00^\circ$& $45.00^\circ$ & $-$ & $45.00^\circ$ & $-$ &$0.00^\circ$& $-82.74^\circ$ \\ \hline
$C_\pm(\ket{\psi^\perp_{5\pi/12}}_3)$  & $45.00^\circ$ & $0.00^\circ$& $45.00^\circ$ & $-$ & $45.00^\circ$ & $-$ &$0.00^\circ$& $-85.75^\circ$ \\ \hline
$C_\pm(\ket{\psi^\perp_{\pi/2}}_1)$  & $45.00^\circ$ & $0.00^\circ$& $45.00^\circ$ & $-$ & $45.00^\circ$ & $-$ &$0.00^\circ$& $-45.00^\circ$ \\ \hline
$C_\pm(\ket{\psi^\perp_{\pi/2}}_2)$  & $45.00^\circ$ & $0.00^\circ$& $45.00^\circ$ & $-$ & $45.00^\circ$ & $-$ &$0.00^\circ$& $-45.00^\circ$ \\ \hline
$C_\pm(\ket{\psi^\perp_{\pi/2}}_3)$  & $45.00^\circ$ & $0.00^\circ$ & $45.00^\circ$ & $-$ & $45.00^\circ$ & $-$ &$0.00^\circ$& $-45.00^\circ$ \\ \hline
$C_\pm(\ket{\psi^\perp_{7\pi/12}}_1)$  & $45.00^\circ$ & $0.00^\circ$ & $45.00^\circ$ & $-$ & $45.00^\circ$ & $-$ &$0.00^\circ$& $-77.92^\circ$ \\ \hline
$C_\pm(\ket{\psi^\perp_{7\pi/12}}_2)$  & $45.00^\circ$ & $0.00^\circ$& $45.00^\circ$ & $-$ & $45.00^\circ$ & $-$ &$0.00^\circ$& $-82.74^\circ$ \\ \hline
$C_\pm(\ket{\psi^\perp_{7\pi/12}}_3)$  & $45.00^\circ$ & $0.00^\circ$& $45.00^\circ$ & $-$& $45.00^\circ$ & $-$ &$0.00^\circ$& $-85.75^\circ$ \\ \hline
$C_\pm(\ket{\psi^\perp_{2\pi/3}}_1)$  & $45.00^\circ$ & $0.00^\circ$ & $45.00^\circ$ & $-$& $45.00^\circ$ & $-$ &$0.00^\circ$& $-69.55^\circ$ \\ \hline
$C_\pm(\ket{\psi^\perp_{2\pi/3}}_2)$  & $45.00^\circ$ & $0.00^\circ$& $45.00^\circ$ & $-$& $45.00^\circ$ & $-$ &$0.00^\circ$& $-76.72^\circ$ \\ \hline
$C_\pm(\ket{\psi^\perp_{2\pi/3}}_3)$  & $45.00^\circ$ & $0.00^\circ$& $45.00^\circ$ & $-$& $45.00^\circ$ & $-$ &$0.00^\circ$& $-81.95^\circ$ \\ \hline
$C_\pm(\ket{\psi^\perp_{3\pi/4}}_1)$  & $45.00^\circ$ & $0.00^\circ$& $45.00^\circ$ & $-$& $45.00^\circ$ & $-$ &$0.00^\circ$& $-64.62^\circ$ \\ \hline
$C_\pm(\ket{\psi^\perp_{3\pi/4}}_2)$  & $45.00^\circ$ & $0.00^\circ$& $45.00^\circ$ & $-$& $45.00^\circ$ & $-$ &$0.00^\circ$& $-72.37^\circ$ \\ \hline
$C_\pm(\ket{\psi^\perp_{3\pi/4}}_3)$  & $45.00^\circ$ & $0.00^\circ$& $45.00^\circ$ & $-$& $45.00^\circ$ & $-$ &$0.00^\circ$& $-78.90^\circ$ \\ \hline
$C_\pm(\ket{\psi^\perp_{5\pi/6}}_1)$  & $45.00^\circ$ &$0.00^\circ$ & $45.00^\circ$ & $-$& $45.00^\circ$ & $-$ &$-24.55^\circ$& $-45.00^\circ$ \\ \hline
$C_\pm(\ket{\psi^\perp_{5\pi/6}}_2)$  & $45.00^\circ$ &$0.00^\circ$ & $45.00^\circ$ & $-$& $45.00^\circ$ & $-$ &$-31.72^\circ$& $-45.00^\circ$ \\ \hline
$C_\pm(\ket{\psi^\perp_{5\pi/6}}_3)$  & $45.00^\circ$ &$0.00^\circ$ & $45.00^\circ$ & $-$& $45.00^\circ$ & $-$ &$36.95^\circ$& $-45.00^\circ$ \\ \hline
$C_\pm(\ket{\psi^\perp_{11\pi/12}}_1)$ & $45.00^\circ$ &$0.00^\circ$& $45.00^\circ$ & $-$& $45.00^\circ$ & $-$ &$-32.93^\circ$& $-45.00^\circ$ \\ \hline
$C_\pm(\ket{\psi^\perp_{11\pi/12}}_2)$  &$45.00^\circ$ & $0.00^\circ$ & $45.00^\circ$ & $-$& $45.00^\circ$ & $-$ &$-37.74^\circ$& $-45.00^\circ$ \\ \hline
$C_\pm(\ket{\psi^\perp_{11\pi/12}}_3)$  & $45.00^\circ$ &$0.00^\circ$& $45.00^\circ$ & $-$& $45.00^\circ$ & $-$ &$-40.75^\circ$& $-45.00^\circ$ \\ \hline
$C_\pm(\ket{\psi^\perp_{\pi}}_1)$  & $45.00^\circ$ & $0.00^\circ$& $45.00^\circ$ & $-$& $45.00^\circ$ & $-$ &$-45.00^\circ$& $-45.00^\circ$ \\ \hline
$C_\pm(\ket{\psi^\perp_{\pi}}_2)$  & $45.00^\circ$ & $0.00^\circ$& $45.00^\circ$ & $-$& $45.00^\circ$ & $-$ &$-45.00^\circ$& $-45.00^\circ$ \\ \hline
$C_\pm(\ket{\psi^\perp_{\pi}}_3)$  & $45.00^\circ$ & $0.00^\circ$& $45.00^\circ$ & $-$& $45.00^\circ$ & $-$ &$-45.00^\circ$& $-45.00^\circ$ \\ \hline\hline
$D(\ket{\psi^\perp_\phi}_{J_x+J_y})$ & $45.00^\circ$ & $45.00^\circ$ & $-$ & $45.00^\circ$ & $-$& $45.00^\circ$ &$0.00^\circ$& $-90.00^\circ$ \\ \hline
\end{tabular}
\label{table:2}
\end{table}

%In our experiment, every term of both sides of inequalities can be obtained directly by the outcomes of the projective measurements. As the previous experiments, the LHS and RHS of the uncertainty inequalities can also be calculated from the density matrices of $\ket{\psi_\phi}$ which are characterized by the quantum state tomography process and the results are shown in Fig.~1. One can see that the results from both methods are in good agreement with theoretical predictions and our direct measurement model works well.

\newpage
\section{verify the new uncertainty relations with a two-level system}

We choose $A=\sigma_x$ and $B=\sigma_y$, and a family of states being measured
$\ket{\psi_\phi}=\sin\phi\ket{+}+\cos\phi\ket{-}$, where $\ket{\pm}$ are the
eigenstates of $\sigma_z=-i\left[\sigma_x,\sigma_y\right]/2$ corresponding to the
eigenvalues $\pm1$. Similar to the case of qutrit, for the special states with
$\phi=\pi/4$ and $\phi=3\pi/4$, the Heisenberg-Robertson relation can be trivial.
Because one of the variances of measurements is zero. The new uncertainty relations
are valid for all the states including those with $\phi=\pi/4$ and $\phi=3\pi/4$.

\begin{figure}
   \includegraphics[width=.5\textwidth]{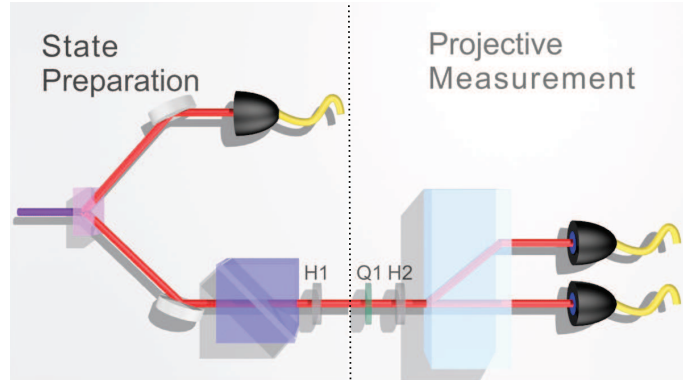}
   \caption{Experimental setup. The herald single photons are created via type-I SPDC in a BBO crystal and are injected into the optical network. The PBS and HWP (H1) are used to generate a qubit state $\ket{\psi_\phi}$. The HWP (or QWP) and the following BD are used to realize the projective measurements of observables $\sigma_{x(y)}$, $\sigma^2_{x(y)}$, $C_\pm$ and $D$.}
\label{setup}
\end{figure}

There is only one state $\ket{\psi^\perp_\phi}=\cos\phi\ket{0}-\sin\phi\ket{1}$
which is orthogonal to the qubit state $\ket{\psi_\phi}$ and it is optimal choice
for the first inequality
\begin{equation}
\Delta A^2+\Delta B^2\geq\pm i\langle\left[A,B\right]\rangle+\left|\bra{\psi}A\pm iB\ket{\psi^\perp}\right|^2,
\label{eq:first}
\end{equation} which becomes an identical equation.
This state is also the only choice for testing the uncertainty inequality in
\begin{equation}
\Delta A^2+\Delta B^2\geq \frac{1}{2}\left|_{A+B}\bra{\psi^\perp}A+B\ket{\psi}\right|^2.
\label{eq:second}
\end{equation}

The observables
\begin{equation}
C_\pm:=(\sigma_x\pm i\sigma_y)\ket{\psi^\perp_\phi}\bra{\psi^\perp_\phi}(\sigma_x\mp i\sigma_y)
\label{eq:C}
\end{equation}and
\begin{equation}
D:=\frac{1}{2}(\sigma_x+\sigma_y)\ket{\psi^\perp_\phi}\bra{\psi_\phi^\perp}(\sigma_x+\sigma_y)
\label{eq:D}
\end{equation} are same as those for the qutrit case in the main
text and for the qubit case, i.e., $J_{x,y,z}=\sigma_{x,y,z}$ and $\ket{\psi^\perp_\phi}_{\sigma_x+\sigma_y}=\ket{\psi^\perp_\phi}$.

\begin{figure}
   \includegraphics[width=.45\textwidth]{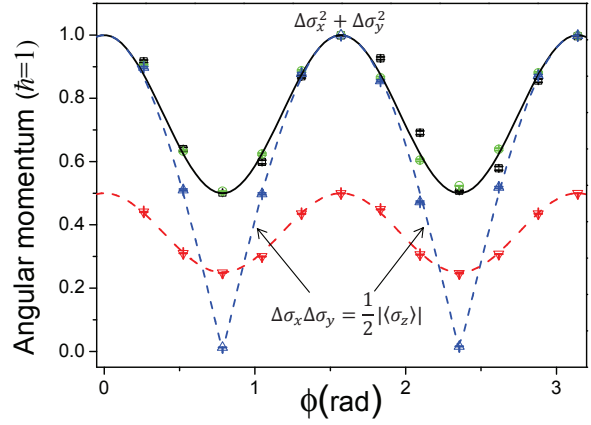}
   \caption{Experimental results.
   The solid black line corresponds to the LHS of the inequalities (\ref{eq:first}) and (\ref{eq:second}). The black squares represent the sum of the measured uncertainties of $\Delta \sigma_x^2$ and $\Delta \sigma_y^2$ with the twelve states $\ket{\psi_\phi}$. The green circles represent the experimental results of the RHS of inequality (\ref{eq:first}) with the optimal and only state $\ket{\psi^\perp_\phi}$ for each of the twelve values of $\phi$. The red dotted lines corresponds to the bound of the inequality (\ref{eq:second}) and red stars represent the measured $\langle D\rangle$ for the twelve states $\ket{\psi^\perp_\phi}$. The blue dotted-dashed curves and triangles represent the theoretical predictions and experimental results of the product of the uncertainties and the expectation value of the commutator (Heisenberg-Robertson relation). Error bars indicate the statistical uncertainty.
   }
\label{data}
\end{figure}

For experimental demonstration shown in Fig.~\ref{setup} of the Appendix,
we prepare a qubit state with the basis states $\ket{0}$ and $\ket{1}$ encoded
in the horizontal and vertical polarizations of heralded single
photons via type I spontaneous parametric down-conversion (SPDC). By changing the
setting angle of the HWP (H1) which applies a rotation on the
polarization qubit, we can obtain the qubit state $\ket{\psi_\phi}$ for
twelve $\phi$. For measurement, a BD and a HWP (or a
QWP) are used to implement measurements of observables $\sigma_x$,
$\sigma_y$, $\sigma_z$, $\sigma^2_x$, $\sigma^2_y$, $C^\perp_\pm$ and $D$. The
setting angles of the HWP and QWP are shown in Table
III of the Appendix.

\begin{table}[htbp]
%\scriptsize
\caption{The setting angles of the Q1 and H2 for the projective measurement stage of two-level system. Here ``$-$" denotes the Q1 is removed from the optical circuit.}
\begin{tabular}{c||c|c}
\hline
Observable & Q1& H2 \\
\hline\hline
$\sigma_x$ & $-$ & $22.50^\circ$ \\ \hline
$\sigma_y$ & $90.00^\circ$ & $22.50^\circ$ \\ \hline
$\sigma^2_x,\sigma^2_y$ & $-$ & $0.00^\circ$  \\ \hline
$\sigma_z$ & $-$ & $0.00^\circ$  \\ \hline\hline
$C_\pm(\ket{\psi^\perp_\phi})$ & $-$ &$0.00^\circ$ \\ \hline
$D(\ket{\psi^\perp_\phi}_{\sigma_x+\sigma_y})$ & $90.00^\circ$ & $\frac{(\phi-\frac{\pi}{2})}{2}$ \\ \hline
\end{tabular}
\label{table:3}

\end{table}

In Fig.~\ref{data} of the Appendix, we show the direct demonstrations of the
new uncertainty relations in (\ref{eq:first}) and (\ref{eq:second}). The solid black line corresponds to
theoretical predictions of the left-hand sides of inequalities (\ref{eq:first}) and (\ref{eq:second}),
i.e., $\Delta\sigma^2_x+\Delta\sigma^2_y$. The black squares represent the sum of
the measured uncertainties ($\Delta\sigma^2_x$ and $\Delta\sigma^2_y$) with the
twelve states. The green circles represent the experimental results of the
right-hand side of inequality (\ref{eq:first}) with the state $\ket{\psi^\perp}$. The red
dotted lines corresponds to the bound of the inequality (\ref{eq:second}) and red stars represent
the measured $\langle D \rangle$ for the twelve states $\ket{\psi^\perp}$. The
blue dotted-dashed curves and triangles represent the theoretical predictions
and experimental results of the product of the uncertainties and the expectation
value of the commutator (the Heisenberg-Robertson relation).

All data of the right-hand side of inequality (\ref{eq:first}) are above the lower
curves which are the product of the uncertainties and the expectation value of the
commutator. Thus the new uncertainty relation in (\ref{eq:first}) is more strengthened compared to the Heisenberg-Robertson relation. Even for the special states with $\phi=\pi/4$ and $\phi=3\pi/4$ which trivialize Heisenberg-Robertson relation, the lower bounds of the new uncertainty inequalities (\ref{eq:first}) and (\ref{eq:second}) are always nontrivial

As we said in the main text, although the new uncertainty relations are valid for
two-level states, however, there is only one state which is orthogonal to the qubit state and it is optimal choice for the first inequality which becomes an identical
equation. Qutrit case displays more features of the new uncertainty relations.

\begin{acknowledgements}
We acknowledge support by NSFC (Nos.~11474049 and 61275122). We thank L. Maccone and A. K. Pati for their helpful discussion.
\end{acknowledgements}


\begin{references}
\bibitem{note} Although the new uncertainty relations are valid for two-level states (seeing more details in the Appendix), however, there is only one state which is orthogonal to the qubit state and it is optimal choice for the first inequality which becomes an identical equation. Qutrit case displays more features of the new uncertainty relations.
\end{references}
\end{document}